\documentclass[final]{siamltex}
\usepackage{amsmath}
\usepackage{mathrsfs}
\usepackage{amsfonts}
\usepackage[pdftex]{graphicx}
\usepackage{psfrag}
\usepackage{amsopn}
\usepackage{amstext}
\usepackage{amssymb}
\usepackage{amscd}
\usepackage{latexsym}

\DeclareMathOperator{\Cov}{Cov}
\DeclareMathOperator{\E}{E}

\DeclareMathOperator{\epi}{epi}
\DeclareMathOperator{\T}{T}

\DeclareMathOperator{\eye}{I}

\DeclareMathOperator{\sign}{sign}

\newcommand{\R}{\mathbb R}

\newcommand{\comment}[1]{}

\DeclareMathOperator{\sig2}{\sigma^2_{max}}
\DeclareMathOperator{\sigg}{\sigma_{max}}

\providecommand{\norm}[1]{\lVert#1\rVert}

\begin{document}

\title{Seven Sins in Portfolio Optimization}

\author{Thomas Schmelzer and Raphael Hauser}

\date{10 October 2013}

\renewcommand{\thefootnote}{\fnsymbol{footnote}}
\renewcommand{\thefootnote}{\arabic{footnote}}

\maketitle

\begin{abstract}
Although modern portfolio theory has been in existence for over 60 years, fund managers often struggle to get its models to produce reliable portfolio allocations without strongly constraining the decision vector by tight bands of strategic allocation targets. The two main root causes to this problem are inadequate parameter estimation and numerical artifacts. When both obstacles are overcome, portfolio models yield excellent allocations. In this paper, which is primarily aimed at practitioners, we discuss the most common mistakes in setting up portfolio models and in solving them algorithmically. 
\end{abstract}

\begin{AMS}
Primary 91G10. Secondary 90C25, 90C90.
\end{AMS}

\begin{keywords} 
Portfolio theory, mean-variance optimization, conic optimization.
\end{keywords}

\section{Introduction}
Modern portfolio theory  \cite{Markowitz, capm, Sharpe} formulates the asset allocation problem as an optimization model, with the objective of maximising the expected portfolio return subject to keeping the estimated risk below a pedefined level (the ``risk budget''). In theory, this approach should result in a  carefully diversified asset allocation across various investable assets. In practice, however, optimal asset allocations computed from portfolio models are often seen as counterintuitive and ill-diversified, as they may contain large positions in just a few assets, and a large number of very small positions \cite{Michaud}. To overcome this problem, fund managers often impose additional constraints in the form of strategic allocation targets. We have found however, that the imposition of allocation targets is artficial and unnecessary if both the parameter estimation and the numerics that go into a portfolio model are carried out very carefully. This paper is aimed primarily at practitioners. Inspired by an essay by Wilmott \cite{WilmottFAQ}, we give a list of seven sins in porfolio optimization that should be avoided at all cost.

\section{The model problem}
The best known investment model is the one-period Mean-Variance (MV) model of Markowitz \cite{Markowitz}. For the purposes of this paper we restrict ourselves to this model, as it is both very simple and widely known among the readership and yet of fundamental interest in finance. A very similar discussion could be held about more complex multiperiod models that involve more general risk terms. For the sake of completeness and to define the notation, we start with a brief recap of the MV model. 

An investor wishes to actively manage a portfolio in $n$ risky
assets. The investor holds fixed positions $x_i(t)$ in asset $i$ over
an investment period $[t,t+1]$ (one hour, day, week, month, ...), at
the end of which he/she is prepared to adjust the positions again.

The expected return of asset $i$ is $\E[R_i]$. $R_i$ is the random variable
describing the return per unit position in asset $i$ over the investment period $[t,t+1]$.
Hence the expected portfolio return is
\[\sum_i x_i\E[R_i] = x^{\T}E[R].\]
The variance of the portfolio return is
\[\sum_{i,j}x_i x_j c_{ij} = x^{\T}\Cov(R,R)x.\]
The elements $c_{ij}$ of the $n \times n$ variance-covariance matrix $\Cov(R,R)$ are
\[
c_{ij} = \Cov\bigl(R_i,R_j\bigr):=\E\left[
\bigl(R_i-\E[R_i]\bigr)\bigl(R_j-\E[R_j]\bigr)
\right].
\]
The problem the investor faces is to decide on the positions $x_i(t)$.
The Mean-Variance approach to this problem proposes to choose
the positions by solving the following optimization problem,
\begin{align}
x(t)=\arg\max_{x\in\R^n}&\,x^{\T}E[R]\label{mean variance}\\
\text{s.t. }&x^{\T}\Cov(R,R)x\leq\sig2,\nonumber
\end{align}
where $\arg\max$ refers to the $x\in\R^n$ where
the objective function is maximised.

To make Model \eqref{mean variance} useable in practice, the
expectations and covariances need to be replaced by estimates
\begin{align*}
\mu_i&\approx\E[R_i],\\
\sigma_{ij}&\approx\Cov\bigl(R_i,R_j\bigr).
\end{align*}
This is usually done via proprietary methods that use historical prices and
other data available at time $t$.

The practical problem being solved is then
\begin{align}
x(t)=\arg\max_{x\in\R^n}&\,x^{\T}\mu,\label{Mean Variance 2}\\
\text{s.t. }&x^{\T}Q x\leq\sig2,\nonumber
\end{align}
where $\cdot^{\T}$ denotes the transpose of a vector,
$\mu$ is the vector of $\mu_i$ and $Q$ is the matrix
of $\sigma_{ij}$.

\begin{figure}[h]
	\centering
		\includegraphics[width=0.75\textwidth]{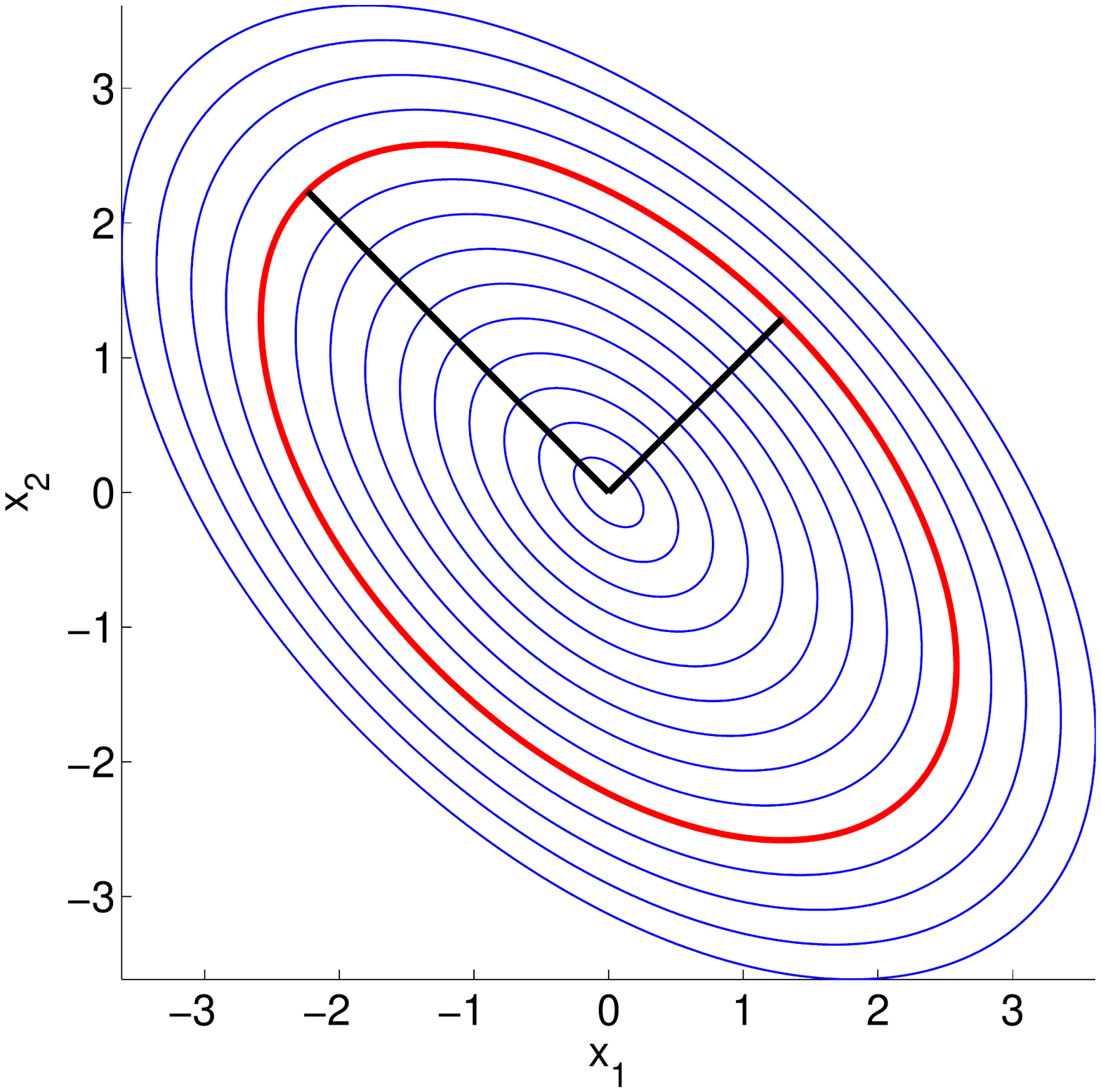}
	\caption[Hans]{Contour lines of the expected variance $x^{\T}Q x$ for $Q =
\begin{pmatrix}
0.2 & 0.1 \\
0.1 & 0.2
\end{pmatrix}$. The red line marks the boundary of the ellipsoid $x^{\T}Q x \leq 1$.  The eigenvectors of $Q$ define the principal directions of the ellipsoid and the inverse of the square roots of the eigenvalues are the corresponding equatorial radii.
}
\label{fig:volCon1}
\end{figure}
The set ${\mathscr F}$
of decision vectors that satisfy the constraints of an optimization problem
is called the {\em feasible domain}. For the model problem this is the interior of an ellipsoid induced by the inequality constraint $x^{\T}Q x\leq\sig2$.

If ${\mathscr F}$ is a convex set (that is, for any pair of points $x,y\in{\mathscr F}$
the line segment $\{\xi x+(1-\xi)y:\,\xi\in[0,1]\}$ between $x$ and $y$
lies in ${\mathscr F}$), and if $f$ is a convex function
(that is, for any $x,y\in{\mathscr F}$ and $\xi\in[0,1]$, $f(\xi x+(1-\xi)y)\leq
\xi f(x)+(1-\xi)f(y)$), then the problem is called a {\em convex optimization
problem}\footnote{An equivalent definition
of a convex problem is to require that the {\em epigraph}
$\epi f:=\{(x,z)\in {\mathscr F}\times\R:\, f(x) \leq \mu\}$ be a convex set.}.

If $Q$ is symmetric positive definite and $\mu \neq 0$, the analytic solution of this problem is given as 
\begin{equation}\label{anaSolution}
x_{*} = \sigg \frac{Q^{-1}\mu}{\sqrt{\mu^{\T}Q^{-1}\mu}}.
\end{equation}
Note that $x_{*}^{\T} Q x_{*} = \sig2$, e.g.~the optimum lies at the boundary of the 
aforementioned ellipsoid. We introduce the Sharpe Ratio \cite{Sharpe} for $x \neq 0$ in this context as
\[
S(x) = \frac{x^{\T}\mu-r_f}{\sqrt{x^{\T}Q x}},
\]
where $r_f$ is a risk-free interest rate over the considered investment period. We will henceforth 
assume short investment periods on the order of days and set $r_f=0$, but all the calculations 
easily extend to the case where $r_f\neq 0$. Note that the Sharpe Ratio $S(x_{*})$ of the optimal 
portfolio does not depend on $\sig2$. 

\section{A List of Seven Sins}
We identified the model formulation, the estimation of parameters and the algorithmic solution as potential mine fields of portfolio optimization.  Another mine field concerns parameter uncertainty. In order to keep the technical difficulty of this paper to a minimum, we chose not to deal with this issue here, but we refer our readers to the extensive recent literature on robust optimization \cite{bertsimas, bertsimasSim, Ceria_Stubbs, ElGhaoui_Oks_Oustry, Goldfarb_Iyengar, gregory,Halldorsson_Tutuncu, hauser-tutuncu, Schottle, Schottle2, Koenig_Tutuncu}. Among the countless ways to produce unsatisfactory results through portfolio modelling, we chose to discuss just seven of the most common mistakes that we have seen committed. 

\subsection{Negative Eigenvalues in the Covariance Matrix}
This is a true classic that deserves the top spot here. A single negative eigenvalue (even if close to zero) can spoil it all.  In practice the covariance matrix $Q$ is estimated using historic data. It is tempting to estimate the elements of the matrix using independent processes for each entry, e.g. moving averages with distinct update rates. This is a recipe for disaster. We illustrate the consequences:

Let $Q$ be a  symmetric real matrix. Then there exists a real orthonormal matrix $V$ such that $D = V^{\T}QV$ is a diagonal matrix. The diagonal entries of $D$ are the eigenvalues of $Q$. The columns of $V$ are the eigenvectors. Let us assume that there exists an eigenvector $v$ associated with a negative eigenvalue $\lambda$. Then $v^{\T}Qv = \lambda v^{\T}v = \lambda < 0$.
Hence we can find a portfolio with a negative variance, that is, $v$ corresponds to a portfolio of negative 
risk. Taking weights $x=\sign(\mu^{\T}v)\tau v$, where $\sign(\mu^{\T}v)=1$ when 
$\mu^{\T}v>0$ and $-1$ otherwise, and where $\tau>0$, we obtain a portfolio with expected return 
$\tau |\mu^{\T}v|$ that satisfies the risk budget, since 
\begin{equation*}
(\tau x)^{\T} Q (\tau x)=-\tau^2\lambda<0<\sig2. 
\end{equation*}
This might entice one to take arbitrarily large positions in the erroneous belief that the risk budget 
will not be exceeded. Some argue that the introduction of extra constraints such as 
the self-financing condition $\sum_i x_i=1$ and non-shortselling conditions $x_i\geq 0$ avoid the 
problem of large positions. However, the positions will still be completely nonsensical, and the 
extra constraints may not be applicable to all asset classes.\footnote{For example, the self-financing condition 
does not make sense for portfolios of future contracts.} Although there exist some approaches to correct estimated and polluted covariances, see Higham\cite{Higham}, we found that a careful analysis of the estimation process itself adds far more value to the trading strategy. 

\subsection{Being Unaware of Ill-Conditioning}
The second sin is of more subtle nature and yet more dramatic.
For almost rank deficient matrices the ellipsoid in Fig. \ref{fig:volCon1} becomes very cigar-shaped. The eigenvectors associated with small eigenvalues will change strongly under even very small perturbations of the matrix. In a typical backtest the entries of such a matrix are updated and hence the ellipsoid will rotate.

Even if $Q$ has only positive eigenvalues they might be still too small.
The inverse of $Q$ is
\[
Q^{-1} = V D^{-1} V^{\T}
\]
where $V$ and $D$ are the aforementioned matrices of eigenvectors and eigenvalues. 
This decomposition can be represented as sum of outer products, that is
\[
Q^{-1} = \sum_{i=1}^n \frac{1}{\lambda_i} v_i v_i^{\T}.
\]
We assume $\lambda_1 \geq \lambda_2 \geq \ldots \geq \lambda_n$. 
Inserting into \eqref{anaSolution} yields
\[
x(t) = \frac{\sigg}{\sqrt{\mu^{\T}Q^{-1}\mu}} \sum_{i=1}^n \frac{v_i^{\T}\mu}{\lambda_i} v_i.
\]
The position $x(t)$ is a linear combination of $n$ \emph{principal portfolios} $v_i$. This term only gives the explicit solution for the standard model. The effect of ill-conditioning is equally important in problems with more constraints. This reveals the fundamental problem of the Markowitz model: 
The weights scale as $\lambda^{-1}$, that is, the optimizer will try to align the position with principal portfolios linked with small eigenvalues. Michaud\cite{Michaud} calls this effect the maximization of the estimation error as the eigenvectors associated with small eigenvalues are most sensitive to noise. In practice, it has often been observed that this results in rather extreme positions.

There are numerous approaches to overcoming this behavior. A popular approach is to ignore eigenvalues smaller than a threshold in the sum for $x(t)$. Often the threshold is induced by the Wigner semicircle distribution. Another powerful approach is to use shrinkage estimators \cite{efron, ledoit1, ledoit2, ledoit3, stein}, the most simple of which is the convex combination
\[
Q' = \kappa Q + (1 - \kappa) \eye, \,\,\, 0 \leq \kappa \leq 1
\]
where $\eye$ is the identity matrix. So the eigenvalues of $Q'$ are $(1-\kappa) + \kappa \lambda_i > (1 - \kappa)$.
Although matrix shrinkage towards $\eye$ does not change the eigenvectors, and hence eigenvectors are still equally sensitive to noise, the eigenvalues change (and are in particular larger than $1 - \kappa$). This weakens the impact of the last few eigenvectors, especially as $\mu$ tends to be close to orthogonal to them.  A third approach that shows excellent practial results was developed by Zuev \cite{Zuev}, who used semi-definite programming to maximize the smallest eigenvalue among the covariance matrices $Q$ that lie in a certain subset of the set of symmetric positive definite matrices.

\subsection{Going the Intermediate Step}
We have observed that some find it simply too tempting to bypass all the maths by using a more trivial intermediate stage in the optimization process. Rather than solving the toy problem \eqref{mean variance},  some practitioners first identify the unit-vector $y$ that results in the maximal expected return. This vector is $\mu/\norm{\mu}$. In a second stage they then identify the scaling factor $\eta$ that achieves 
to achieve $\eta^2 y^{T}Qy = \sig2$. Using this approach the optimal vector is
\[
y^{*} = \eta y = \sigg \frac{\mu}{\sqrt{\mu^T Q \mu}}.
\]
Although this looks close to the analytic solution \eqref{anaSolution}, dramatic differences exist: In the numerator of $y^{*}$ all information contained in $Q$ is ignored. As a result, this approach does not exploit diversification, one of the uncontested benefits of investing in a portfolio rather than a single asset. 
Worse even, there is some danger lurking in the denominator. As soon as $\mu$ is aligned with an eigenvector corresponding to  a small eigenvalue the denominator can be extremely small and blow up the position size and hence the portfolio.
 
A modern solver may iterate through the feasible domain by taking dozens of intermediate steps that converge towards the unique global maximum of \eqref{Mean Variance 2}. However, the choice of those steps should be left to the solver.  

\subsection{Failing to Recognize Convexity}
Most portfolio optimization models are convex programming problems, a class of well-studied optimization problems with rich mathematical structure \cite{Ben-Tal, boyd}. Non-convex optimization problems may have multiple local extrema and convergence to a global extrema (it may be not unique) cannot be guaranteed as the solver may get attracted  to any of the local extrema \cite{nocedal}. When multiple optimization problems are solved sequentially, the solution may get attracted to a different local extremum each time, even when the model parameters change only slightly, and this can result in high trading costs. It is therefore often preferrable to use a convex portfolio model that avoids these costs or at least do extensive backtesting to estimate the impact of these costs on performance. 

A related problem is the failure to recognize convexity when it is present implicitly in a model, as this 
prevents one from using convex optimization software that would be able to solve the problem much 
efficiently and robustly than most nonconvex solvers. As an example, consider a portfolio manager who 
aims to maximize the Sharpe Ratio \cite{Sharpe} 
\begin{equation}
x(t)=\arg\max_{x\in\R^n}\,S(x).\label{Sharpe 2}
\end{equation}
We note that the Sharpe Ratio is not defined for $x \neq 0$ and 
the limit of $S(x)$ for $\norm{x}\rightarrow 0$ does not exist. Further, the Sharpe Ratio function has no 
unique maximizer, since $S(\tau x) = S(x)$ for all $\tau > 0$ and $x \neq 0$. 

Using model \eqref{Sharpe 2} 
in conjunction with a nonconvex solver may lead to the evaluation of $S(x)$ for points close to the origin, 
which causes numerical inaccuracy. To avoid this problem, we wish to avoid the origin and a reasonably 
large region around it. Exploiting the fact that $S(\tau x) = S(x)$, we may rescale $x$ to satisfy  $\sqrt{x^{\T}Qx}=\sigg$. Hence, the problem becomes
\begin{align}
x(t)=\arg\max_{x\in\R^n}&\,S(x),\nonumber\\
\text{s.t. }&\sqrt{x^{\T}Q x} = \sigg,\nonumber.
\end{align}
The numerator of $S(x)$ now being constrained to a fixed number, it plays no role in the maximization 
of $S(x)$ and can be neglected, that is, the optimal decision vector $x(t)$ can be found by solving the 
simpler problem\footnote{Recall that we assumed the risk free rate to be zero.}  
\begin{align}
x(t)=\arg\max_{x\in\R^n}&\,x^{\T}\mu,\nonumber\\
\text{s.t. }&x^{\T}Q x = \sig2,\label{the rescaling constraint}
\end{align}
This problem is still not convex, as the feasible domain is the boundary of the risk ellipsoid, but it 
provably has the same solution as the convex problem 
\begin{align}
x(t)=\arg\max_{x\in\R^n}&\,x^{\T}\mu,\label{Sharpe new}\\
\text{s.t. }&x^{\T}Q x \leq\sig2,\nonumber
\end{align}
for generically we have $\mu^{\T}x(t)>0$, and if $x^{\T}(t)Q x(t)<\sig2$, then $\tau x(t)$ is 
an improved feasible solution for $\tau=\sigg/(x^{\T}(t)Q x(t))^{1/2}>1$. This contradicts the 
optimality of $x(t)$ and shows that the optimal $x(t)$ must satisfy the constraint \eqref{the rescaling constraint} of the nonconvex problem automatically. In this fashion, solving the convex problem 
\eqref{Sharpe new} yields a maximizer $x(t)$ of the nonconvex problem \eqref{Sharpe 2} whilst 
avoiding the associated numerical problems. 

\subsection{Using the Wrong Solver}
Most portfolio managers solve a variant of \eqref{mean variance} with additional constraints for which the 
optimal solution is no longer given in closed form. A solution has to be computed algorithmically in this case, 
and the choice of a good solver is crucial. The most robust and powerful solvers require the problem to be 
reformulated in a specific standard form, which usually requires a lift-and-project approach that will be 
further described below. For example, Matlab's \texttt{quadprog} is unable to solve problem 
\eqref{Mean Variance 2} directly without lifting the risk constraint into to the utility function. 

Bypassing such steps by writing proprietary optimization software is not recommended, as the numerical challenges faced in such attempts are easily underestimated. A particularly popular route in this context is to 
use \emph{simulated annealing}, because it is intuitive to understand and trivial to code. However, 
whilst simulated annealing has its place in highly nonconvex and unstructured global optimization problems, 
it is not designed to exploit any of the strong mathematical structures that underly portfolio problems, in 
particular convexity, and as a result it converges far less robustly and exceedingly more slowly than modern 
convex optimization solvers. 

To demonstrate the fallacy of simulated annealing we tried to implement a simple scheme. However, even the most basic convex and constrained problems are hard to approximate with such algorithms and require careful tuning. It remains a black art. In a multi-period setting in which the portfolio positions are frequently updated, it is of utmost importance to choose a robust solver that yields reproducible results free of random fluctuations that cause unnecessary transactions costs when rebalancing the portfolio. The issue of transaction costs is so important that is wise to reduce the fluctuations that occur as a function of the changing model parameters even further, by ways of regularization terms. The use of a fast, robust, reliable, deterministic algorithm is also hugely important in any back-testing framework for quantitative trading in which one wishes to study the statistical behavior of the trading strategy and avoid all avoidable sources of artificial randomness.  

Most portfolio problems have reformulations as second order cone programming problems (SOCPs), 
or, occasionally semidefinite programming problems (SDPs). Such problems can be solved in 
polynomial time via iterative schemes known as  {\em interior-point methods}. Practical implementations 
of such schemes are currently the leading codes for solving portfolio problems in terms of robustness, 
speed and reliability. Leading codes include the following:

\begin{itemize}
\item SDPT3 \cite{sdpt3}, see {\ttfamily http://www.math.nus.edu.sg/$\sim$mattohkc/sdpt3.html},  
is freely available, only available for Matlab.
\item SeDuMi \cite{sedumi}, see {\ttfamily http://sedumi.ie.lehigh.edu/}, is very similar to SDPT3. 
\item MOSEK \cite{MarkowitzMosek}, see {\ttfamily http://www.mosek.com} is the leading commercial conic 
programming solver and has interfaces for the most common programming languages.
\end{itemize}

\subsection{Ignoring the Lift}
Lifting is a powerful technique to make hard problems look simple by introducing additional dimensions.
Using a classic example we demonstrate what it is all about: Often problems of type \eqref{Mean Variance 2} are solved in a large loop over thousands of investment periods. For each period some of the estimates may change and the portfolio is rebalanced. It is certainly a good idea to take into account some costs when changing the position. Most quants use quadratic costs as the resulting cost term is differentiable. However, such terms tend to overestimate costs for large trades.

Let us assume that the system currently holds the position $x^0$ and a new investment periods starts. The formulation above is not taking into account the position $x^0$ at all. This may result in large costs and spurious oscillations in the position.To address these problems we penalize deviations from $x^0$, that is
\begin{align}
x(t) = \arg\max_{x\in\R^{n}}&\,x^{\T} {\mu} - \sum_{i=1}^n p_i \lvert x_i - x^0_i \rvert \label{LTTF1} \\
\text{s.t. }&x^{\T}C x\leq\sig2. \nonumber
\end{align}
The positive parameters $p_i$ reflect the estimate costs per unit position.
Introducing auxiliary variables $t_i$ the problem can be reformulated as
\begin{align}
x(t) = \arg\max_{(x, t)\in\R^{2n}}&\,\left[x^{\T} {\mu} - \sum_{i=1}^n p_i t_i \right] \label{LTTF2}\\
\text{s.t. }&x^{\T}C x\leq\sig2, \nonumber \\
& x_i - x^0_i \leq t_i,\nonumber \\
& x^0_i - x_i \leq t_i.\nonumber
\end{align}

Problem \eqref{LTTF2} is called a {\em lifting} of Problem
\eqref{LTTF1}. Through the introduction of extra variables, liftings
involve an inflation of the problem dimension. This seeming disadvantage
is offset when the following occurs:
\begin{itemize}
\item The lifted problem belongs to a problem class with lower
complexity than the original problem. For example, in the above case,
a nonsmooth problem turned into a smooth one (the nondifferentiable
absolute value terms in the constraint have disappeared). In other cases,
nonconvex problems can be convexified through liftings.
\item The lifted problem belongs to a problem class for which efficient standard
software already exists, avoiding the need to implement a custom designed
algorithm. In the example above, a quadratic programm resulted, one of the most efficiently solved
class of optimization problems.
\end{itemize}

Using more advanced conic techniques, one can also model more realistic cost models such as
$\sum_{i=1}^n p_i \lvert x_i - x^0_i\rvert^{3/2}$ or $\sum_{i=1}^n p_i \lvert x_i - x^0_i\rvert^{4/3}$.

\subsection{Solving the Impossible}
The reason why sometimes even the most sophisticated solver cannot find a solution is because there is none.
A feasible domain can be empty. A modern solver can detect such a situation.

As long as $\sig2 \geq 0$ problem \eqref{Mean Variance 2} has a non-empty feasible domain as it contains the trivial portfolio $x = 0$.
In equity portfolios the position $x$ is often interpreted as a fraction of the investors's capital. Being \emph{fully invested} is then reflected by the constraint
$\sum x_i = e^{\T} x = 1$.
The enhanced problem
\begin{align}
x(t)=\arg\max_{x\in\R^n}&\,x^{\T}E[R]\label{equity}\\
\text{s.t. }&x^{\T}\Cov(R,R)x\leq\sig2,\nonumber\\
&e^{\T} x = 1 \nonumber
\end{align}
has no solution if $\sig2$ is smaller than the minimum variance $\sigma^{2}_{*}$ of all fully invested portfolios, e.g.
\begin{align}
\sigma^{2}_{*}=\min_{x\in\R^n}&\,x^{\T}\Cov(R,R)x\label{equity2}\\
\text{s.t. }&e^{\T} x = 1. \nonumber
\end{align}


\begin{thebibliography}{29}

\bibitem{MarkowitzMosek}
E.D.\ Andersen, J.\ Dahl and H.A.\ Friberg. 
Markowitz portfolio opitmization using MOSEK. 
{\em MOSEK Technical report TR-2009-2}. 


\bibitem{Ben-Tal}
A.\ Ben-Tal and A.\ Nemirovski.
{\em Lectures on Modern Convex Optimization: Analysis, 
Algorithms, and Engineering Applications}. 
MPS-SIAM Series on Optimization, SIAM, 
Philadelphia, PA, 2001. 

\bibitem{bertsimas}
D.\ Bertsimas, D.B.\ Brown and C.\ Caramanis. 
Theory and applications of robust optimization. 
{\em SIAM Rev.}.
Vol.\ 53, no.\ 3, pp.\ 464--501, 2010. 

\bibitem{bertsimasSim}
D.\ Bertsimas and M.\ Sim. 
The price of robustness. 
{\em Oper.\ Res.}. 
Vol.\ 52, no.\ 1, pp.\ 35--53, 2004. 

\bibitem{boyd}
S.\ Boyd and L.\ Vandenberghe. 
Convex optimization. 
{\em Cambridge University Press}, Cambridge, 2004.

\bibitem{Ceria_Stubbs}
S.\ Ceria and R.\ Stubbs.
Incorporating estimation errors into portfolio selection: robust portfolio construction. 
{\em Axioma Research Paper}, no.\ 003, May 2006. 

\bibitem{efron}
B.\ Efron. 
Biased versus unbiased estimation. 
{\em Adv.\ Math.},
Vol.\ 16, pp.\ 259–277, 1975.

\bibitem{ElGhaoui_Oks_Oustry} 
L.\ El Ghaoui, M.\ Oks, and F.\ Lebret.
Worst-case value at risk and robust portfolio optimization: A conic programming approach. 
{\em Operations Research}. 
Vol.\ 51, no.\ 4, pp. 543--556, 2003.

\bibitem{Goldfarb_Iyengar}
D.\ Goldfarb and G.\ Iyengar. 
Robust Portfolio Selection Problems. 
{\em Mathematics of Operations Research}, 
Vol. 28, no. 1, pp. 1--38, 2003.

\bibitem{gregory}
C.\ Gregory, K.\ Darby-Dowmann and G.\ Mitra. 
Robust optimization and portfolio selection: The cost of robustness. 
{\em European Journal of Operations Research}. 
Vol.\ 212, pp.\ 417--426.
 
\bibitem{Halldorsson_Tutuncu}
B.V.\ Halld\'{o}rsson and R.H.\ T\"{u}t\"{u}nc\"{u}.
An interior-point method for a class of saddle point problems.
{\em Journal of Optimization Theory and Applications}. 
Vol.\ 116, no.\ 3, pp.\ 559--590, 2003.

\bibitem{hauser-tutuncu}
R.A.\ Hauser, V.\ Krishnamurthy and R.\ T\"ut\"unc\"u. 
Relative Robust Portfolio Optimization. 
arXiv:1305.0144. 

\bibitem{Higham}
N.J.\ Higham. 
Computing the nearest correlation matrix -- a problem from finance. 
{\em IMA J.\ Numer.\ Anal.}, 
22, 2002(3), pp.\ 329 -- 343.

\bibitem{ledoit1}
O.\ Ledoit and M. Wolf.
Improved estimation of the covariance matrix of stock returns with an application to portfolio selection. 
{\em J. Empir. Finance}.
Vol.\, 10, pp.\ 603–621, 2003.

\bibitem{ledoit2}
O.\ Ledoit and M. Wolf.
Honey, I shrunk the sample covariance matrix. 
{J.\ Portfolio Management}.
Vol.\ 30, pp.\ 110–119, 2004.

\bibitem{ledoit3}
O.\ Ledoit and M. Wolf. 
A well conditioned estimator for largedimensional covariance matrices. 
{\em J.\ Multiv.\ Anal.} 
Vol.\ 88, pp.\ 365–411, 2004.

\bibitem{Markowitz}
H.M.\ Markowitz. 
Portfolio Selection.
{\em Journal of Finance}. 
Vol.\ 7, pp.\ 77--91, 1952. 

\bibitem{Michaud}
R.O.\ Michaud. 
{\em Efficient Asset Management}, 
Wiley, 1st edition, 1998.

\bibitem{nocedal} 
J.\ Nocedal and S.\ Wright. 
Numerical optimization. Second edition. 
{\em Springer Series in Operations Research and Financial Engineering}.  
Springer, New York, 2006. 

\bibitem{Schottle} 
K.\ Sch\"ottle, R.\ Werner and R.\ Zagst. 
Comparison and robustification of Bayes and Black-Litterman models. 
{\em Math.\ Methods of Oper.\ Res.}. 
Vol.\ 71, no.\ 3, pp.\ 453--475, 2010. 

\bibitem{Schottle2}
K.\ Sch\"ottle and R.\ Werner. 
Robustness properties of mean-variance portfolios. 
{\em Optimization}. 
Vol.\ 58, no.\ 6, pp.\ 641--663, 2009. 

\bibitem{capm} 
W.F.\ Sharpe. 
Capital asset prices: A theory of market equilibrium under conditions of risk. 
{\em Journal of Finance}. 
Vol.\ 19, no.\ 3, pp.\ 425-–442, 1964.

\bibitem{Sharpe} 
W.F.\ Sharpe.
The Sharpe Ratio. 
{\em Journal of Portfolio Management}. 
Vol.\ 21, no.\ 1, pp.\ 49--59, 1994. 

\bibitem{sedumi}
J.\ Sturm and I.\ Polik. 
SeDuMi 1.05 R5 user's guide. 
{\em http://sedumi.ie.lehigh.edu}. 

\bibitem{stein}
C.\ Stein. 
Inadmissibility of the usual estimator for the mean of a multivariate distribution. 
In J. Neyman (Ed.), {\em Proc.\ Third.\ Berkeley Symp.\ Math.\ Statist.\ Probab.}, 
Vol.\ 1, pp.\ 197–206. Univ. California Press, 1956.


\bibitem{sdpt3} 
M.J.\ Todd, K.C.\ Toh and R.H.\ T\"ut\"unc\"u. 
On the implementation and usage of SDPT3 – a Matlab software package for semideﬁnite-quadratic-linear programming, version 4.0. 
{\em http://www.math.nus.edu.sg/~mattohkc/sdpt3.html}. 

\bibitem{Koenig_Tutuncu}
R.H.\ T\"{u}t\"{u}nc\"{u} and M.\ Koenig.
\newblock Robust Asset Allocation.
{\em Annals of Operations Research},
Vol. 132, pp. 157--187, 2004.

\bibitem{WilmottFAQ} 
P.\ Wilmott.
The Commonest Mistakes in Quantitative Finance: A Dozen Basic Lessons in Commonsense for Quants and Risk Managers and the Traders Who Rely on Them. 
{\em Frequently Asked Questions in Quantitative Finance, 2nd edition}, 
Wiley, 2009, pp.\ 313--382.

\bibitem{Zuev}
D.\ Zuev.
{\em New and Improved Robust Portfolio Selection Methods}, DPhil.\ Thesis, Mathematical Institute, University of Oxford, 2009.


\end{thebibliography}
\end{document}